\documentclass{PoS}
\usepackage[utf8]{inputenc}
\usepackage{amsmath,amssymb}
\usepackage{graphicx}
\usepackage{enumerate}
\usepackage{multirow}
\usepackage[square,numbers]{natbib}
\usepackage[squaren,Gray]{SIunits}

\def\plots{0.8}
\graphicspath{{figures/}}

\newcommand{\bigo}{\mathrm{O}}
\newcommand{\cf}{\textit{cf.}~}
\newcommand{\err}[2]{(#1)_{\mathrm{#2}}}
\newcommand{\ev}{\electronvolt}
\newcommand{\fm}{\femto\meter}
\newcommand{\mQCD}{\mathrm{QCD}}
\newcommand{\SU}{\mathrm{SU}}
\newcommand{\U}{\mathrm{U}}

\title{Systematic errors in partially-quenched QCD plus QED
  lattice simulations}

\ShortTitle{Systematic errors in partially-quenched QCD plus QED
  lattice simulations}

\author{\speaker{A. Portelli}$\,^{b}$, S. Dürr$\,^{a,d}$,             
        Z. Fodor$\,^{a,c,d}$, J. Frison$\,^{b}$, C. Hoelbling$\,^{a}$,
        S.D. Katz$\,^{a,c}$, \mbox{S. Krieg}$\,^{a,d}$, T. Kurth$\,^{a}$, 
        L. Lellouch$\,^{b}$, T. Lippert$\,^{a}$, 
        A. Ramos$\,^{b}$ and K.K. Szabó$\,^{a}$ 
        \mbox{(Budapest-Marseille-Wuppertal Collaboration)}\\
        $\,^{a}$Bergische Universität Wuppertal, 
        Gau\ss str. 20, D-42097 Wuppertal, Germany\\
        $\,^{b}$Centre de Physique Théorique, Aix-Marseille Univ,
        CNRS UMR 7332, Univ Sud Toulon Var, 13288 Marseille cedex 9, France
        \thanks{Unité Mixte de Recherche (UMR 7332) du CNRS, de l'Université
        d'Aix-Marseille et de l'Université Sud Toulon Var. Unité affiliée à la    
        FRUMAM, Fédération de Recherche 2291 du CNRS.}\\
        $\,^{c}$Institute for Theoretical Physics, Eötvös University,
        H-1117 Budapest, Hungary\\
        $\,^{d}$Jülich Supercomputing Centre, Forschungszentrum Jülich,
        D-52425 Jülich, Germany\\
        ~\\
        E-mail : \email{portelli@cpt.univ-mrs.fr}}

\abstract{At the precision reached in current lattice QCD
  calculations, electromagnetic effects are becoming numerically
  relevant. Here, electromagnetic effects are included by
  superimposing $\U(1)$ degrees of freedom on $N_f=2+1$ QCD
  configurations from the Budapest-Marseille-Wuppertal
  Collaboration. We present preliminary results for the electromagnetic
  corrections to light pseudoscalars mesons masses and discuss some of
  the associated systematic errors.}

\FullConference{The XXIX International Symposium on Lattice Field Theory - 
                Lattice 2011\\
                July 10-16, 2011\\
                Squaw Valley, Lake Tahoe, California}

\begin{document}

\section{Motivation}
Isospin is a near symmetry of the hadron spectrum because it is only
broken by small effects: 
\begin{enumerate}[(i)]
\item the mass difference $m_{u}-m_{d}$
\item the difference in the charge of the $u$ and the $d$ quark
\end{enumerate}
which are summarized in the following table\footnote{We have chosen
  to quote the quark masses from
  \citep{Durr:2010uf,Durr:2010tu}. Please see \citep{Colangelo:2010uu}
  for a complete list of lattice determinations of these masses.}:
\begin{center}
\begin{tabular}{|l|c|c|}
\cline{2-3}
\multicolumn{1}{c|}{~} & $u$ & $d$\\
\hline
Mass ($\mega\ev$) \citep{Durr:2010uf,Durr:2010tu} & $2.15\err{03}{stat.}\err{10}{sys.}$ & 
$4.79\err{07}{stat.}\err{12}{sys.}$\\
Charge & $\frac{2}{3}e$ & $-\frac{1}{3}e$\\	
\hline
\end{tabular}
\end{center}

\noindent These effects are expected to be at the percent level. The size of 
mass breaking is the mass difference $m_{u}-m_{d}$ relatively to a typical QCD 
scale $\Lambda_{\mQCD}$ and the order of electromagnetic breaking is the fine 
structure constant at zero momentum 
$\alpha=\frac{e^{2}}{4\pi}\simeq\frac{1}{137}$. However, their importance is
not commensurate to their size. For instance, they are responsible for the
stability of matter through the proton-neutron mass difference. Moreover, 
lattice QCD calculations have recently approached percent or even sub-percent 
precision \cite{Hoelbling:2011kk,Mawhinney:2011}, so the inclusion of these 
effects becomes relevant.

Another interesting isospin breaking quantity is the \emph{absolute correction 
to Dashen's theorem}
\begin{equation}
\Delta_{A}D=\Delta_{\mathrm{EM}}M_{K}^{2}-\Delta_{\mathrm{EM}}M_{\pi}^{2}
\end{equation}
where :
\begin{equation}
\Delta_{\mathrm{EM}}M_{P}^{2}=(M_{P^{+}}^{2}-M_{P^{0}}^{2})_{m_{u}=m_{d}}
\end{equation}
is the \emph{electromagnetic squared mass splitting} of the isospin multiplet 
$P$. One can also consider the dimensionless \emph{relative correction to 
Dashen's theorem} :
\begin{equation}
\Delta_{R}D=
\frac{\Delta_{\mathrm{EM}}M_{K}^{2}}{\Delta_{\mathrm{EM}}M_{\pi}^{2}}-1
\end{equation}
R. Dashen has shown in~\citep{Dashen:1969tn} than $\Delta_{A}D=0$ in the 
$\SU(3)$ chiral limit and that the leading corrections are $\bigo(\alpha 
m_{s},\alpha^{2})$. The quantity $\Delta_{A}D$ is interesting because it is 
very sensitive to the up and down quark masses. Moreover, a precise estimation 
of these quantities has not been made until now (\cf Table \ref{tab:dash}).

\begin{table}[h!]
\centering
\begin{tabular}{|r|c|c|l}
    \cline{2-3}
    \multicolumn{1}{c|}{~} & $\Delta_{A}D~(\mega\ev\squared)$ & 
    $\Delta_{R}D$ & \\
    \cline{1-3}
    \multirow{6}*{\rotatebox{90}{phenomenology}}
    & $1230$                & $\mathbf{0.80}$           &
    \citet{Donoghue:1993bm}~(\citeyear{Donoghue:1993bm})    \\
    & $\mathbf{1300(400)}$  & $1.0(3)$                  &
    \citet{Bijnens:1993go}~(\citeyear{Bijnens:1993go}) \\
    & $\mathbf{360}$        & $0.26$                    &
    \citet{Baur:1996gf}~(\citeyear{Baur:1996gf})        \\
    & $\mathbf{1060(320)}$  & $0.9(4)$                  &
    \citet{Bijnens:1997ku}~(\citeyear{Bijnens:1997ku})     \\
    & $\mathbf{1080}$       & $\mathbf{0.68}$           &
    \citet{Gao:1997hq}~(\citeyear{Gao:1997hq})         \\
    & $\mathbf{1070}$       & $0.74$                    &
    \citet{Bijnens:2007jk}~(\citeyear{Bijnens:2007jk})     \\
    \cline{1-3}
    \multirow{3}*{\rotatebox{90}{lattice~~~~~~~~~~}}
    & $526$                 & $0.39$                    &
    \citet{Duncan:1996cy}~(\citeyear{Duncan:1996cy})      \\
    & $\mathbf{340(92)}$    & $0.30(8)$                 &
    \citet{Blum:2007gh}~(\citeyear{Blum:2007gh})         \\
    & $\mathbf{1250(550)}$  & N/A                       &
    \citet{Basak:2008td}~(\citeyear{Basak:2008td})        \\
    & $\mathbf{830(180)}$   & $\mathbf{0.60(14)}$       & 
    \citet{Portelli:2010tz}~(\citeyear{Portelli:2010tz})     \\
    & 707(75)               & $\mathbf{0.63(6)}$ &
    \citet{Blum:2010ym}~(\citeyear{Blum:2010ym})         \\
    \cline{1-3}
\end{tabular}
\caption{Results for the violations to Dashen's theorem in phenomenelogy and 
lattice computations. Order is chronological. Bold numbers are the results
given by the authors, the others are deduced from information given in the 
corresponding paper.
\label{tab:dash}}
\end{table}

\section{Simulation setup}
For this work, we used a subset of our 2010 $\SU(3)$ gauge
configurations~\citep{Durr:2010tu}. These configurations were
generated using $N_f=2+1$ QCD simulations with the tree-level Symanzik
gauge action, tree level $\bigo(a)$-improved Wilson fermions and two
steps of HEX smearing. We have already used this dataset to compute light
quark masses~\citep{Durr:2010uf,Durr:2010tu} and the kaon bag
parameter~\citep{Durr:2011mp}.

To include QED, we generate real electromagnetic fields $A_{\mu}$
using a non-compact formulation (\cf \citep{Portelli:2010tz} for more
details). Then we phase $\SU(3)$ strong links by the $\U(1)$ links
$\exp(iQA_{\mu})$, where $Q$ is a chosen electric charge. The resulting
$\U(3)$ links are used inside the Dirac-Wilson operator to compute
quark propagators. This method allows to use previously generated
$\SU(3)$ gauge configurations to obtain results, but the
electromagnetic vacuum polarization is not taken into account. For the
preliminary study presented here, the
masses of the up and down valence quarks are tuned
individually such that the bare masses are equal. This occurs when
these masses become equal to the light sea quark
mass~\citep{Portelli:2010tz}. The $\pi^0$ squared mass is obtained by
averaging squared ground state energies obtained with $\bar{u}u$ and
$\bar{d}d$ connected pseudoscalar correlators.  This is correct up to
NLO isospin breaking corrections.

In this setup, one is confronted with two new types of systematics
effects : the quenching of the electromagnetic field and
electromagnetic finite volume effects.

\section{QED quenching errors}
Using photon fields without vaccuum polarization leads to partial
quenching effects: valence quarks have electric charges but sea quarks
do not. These effects can be evaluated in partially quenched
chiral perturbation theory with QED (PQ$\chi$PT+QED). Next to leading
order (NLO) $\SU(3)$ formulas for self-energies and decay constants
can be found in~\citep{Bijnens:2007jk,Blum:2010ym}. The $\SU(3)$ NLO
sea contribution to a pseudoscalar meson squared mass is given by :
\begin{equation}\label{eq:qeffect}
\delta_{\mathrm{sea}}M_{ij}^{2}=
\frac{e^{2}C}{8\pi^{2}F_{0}^{4}}(q_{i}-q_{j})
\sum_{s=4}^{6}q_{s}\left[\chi_{is}\log\left(\frac{\chi_{is}}{\mu}\right)
                    -\chi_{js}\log\left(\frac{\chi_{js}}{\mu}\right)\right]
-\frac{e^{2}}{3}Y_{1}\chi_{ij}\sum_{s=4}^{6}q_{s}^{2}
\end{equation}
where $i$ and $j$ are the valence indices of the quarks composing the 
meson, indices, $s$, between $4$ and $6$ are sea quark indices, 
$\chi_{kl}=B_0(m_k+m_l)$, $e\simeq 0.302822$ is the positron electric charge, 
$q_k$ are quark charges in units of $e$ and $Y_1$, $C$ and $F_0$ are low energy 
constants (LECs).

For the results presented here, the sea and valence masses are very
nearly equal and we can assume that the absence of sea charges is the
only partial quenching effect. In that case, one can check easily that
(\ref{eq:qeffect}) is independant of the scale $\mu$. The LECs $C$ and
$F_0$ are essentially known~\citep{Bijnens:2007jk}, but $Y_1$, which
is a sea electromagnetic contribution, is unknown. RBC, who studied
other partially-quenched QED LECs~\citep{Blum:2007gh,Blum:2010ym},
considers that it would be unnatural that $Y_1>10^{-2}$. Thus, in the
following, we will use $Y_1=10^{-2}$. With
(\ref{eq:qeffect}), one obtains the following partial quenching
error estimates:
\begin{itemize}
	\item negligible (less than $0.1\%$) for 
	$M_{K^{+}}$,~$M_{K^{0}}$,~$\Delta_{\mathrm{EM}}M_{\pi}$ and
	$\Delta_{\mathrm{EM}}M_{\pi}^{2}$
	\item $\sim 0.4\%$ for $M_{\pi^{0}}$
	\item $\sim 5\%$ for $\Delta_{\mathrm{EM}}M_{K}$ and 
	$\Delta_{\mathrm{EM}}M_{K}^{2}$
\end{itemize}

\section{Electromagnetic finite volume effects}
It is already known~\citep[p. 18]{Durr:2010tu} that the Budapest-Marseille-Wuppertal 
gauge ensembles have spatial volumes that allow to neglect QCD finite 
volume effects. However, as electromagnetism is a long range interaction, one 
might expect large QED finite volume effects on a periodic lattice. 
Predictions for these effects have been made in 
PQ$\chi$PT+QED~\citep{Hayakawa:2008ci,Blum:2010ym}. These formulas are far 
from simple and complicate chiral fits. Additionally, their 
predictivity seems limited~\citep[p. 16]{Blum:2010ym}.

\begin{figure}[h]
\centering
\includegraphics[width=\plots\textwidth]{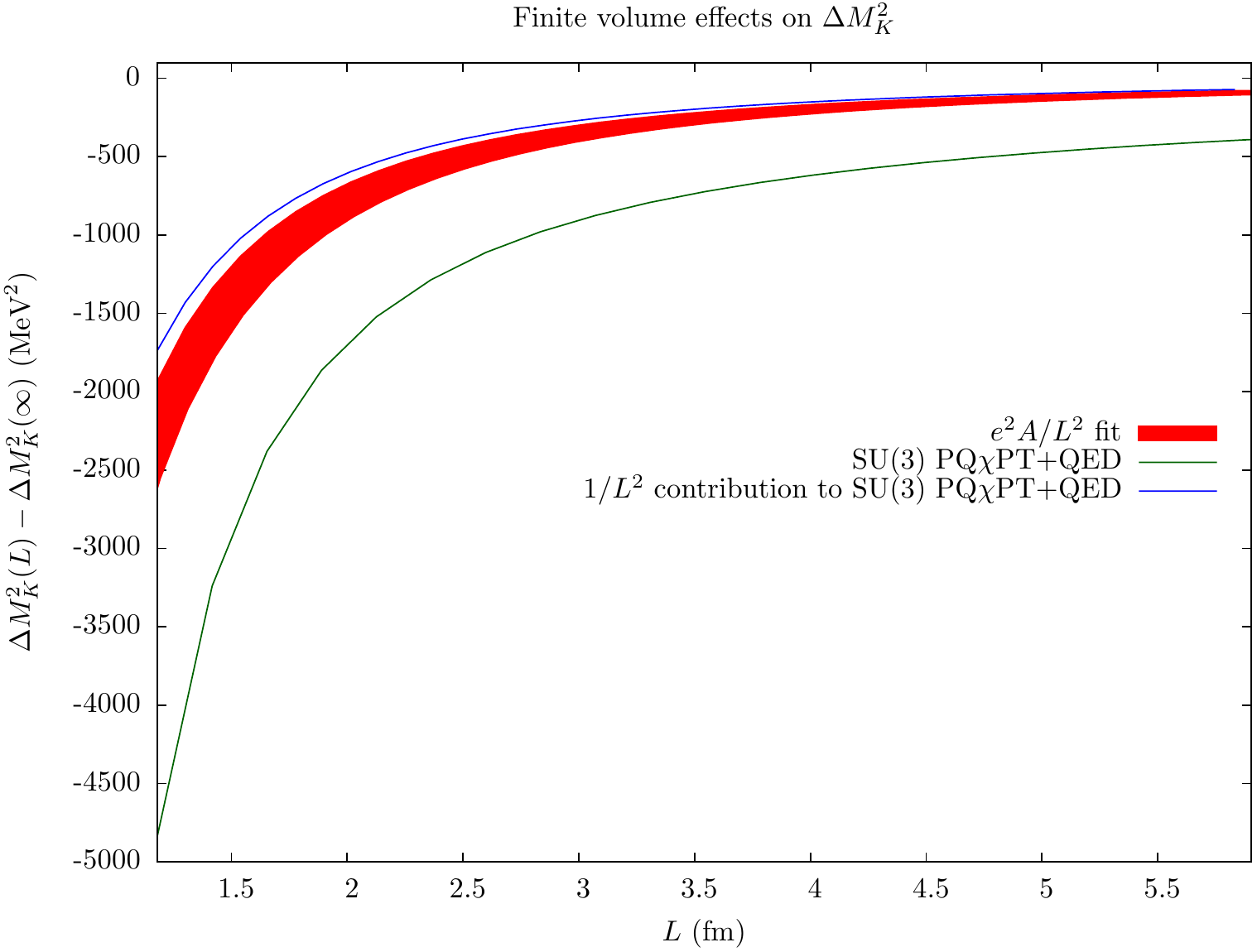}
\caption{Comparison of our results with different models of finite
  volume corrections for the kaon squared mass splitting, as a
  function of lattice spatial extent. The red band represents the fit
  of our lattice data to the model (\protect\ref{eq:fvol}), with
  $M_{\pi^0}$ fixed to its physical value. Only statistical errors are
  shown.  The green curve represents the prediction of $\SU(3)$
  PQ$\chi$PT+QED, which depends only on the LO, LECs $F_0$ and $C$,
  given in \cite{Bijnens:2007jk}. There is a clear disagreement
  between the two. However, the $\frac{1}{L^2}$ contribution
  from the full $\SU(3)$ PQ$\chi$PT+QED expression (blue curve)
  roughly agrees with our data.
\label{fig:fvol}}
\end{figure}

Here we choose a more straightforward approach. Dimensional analysis 
suggests that the leading finite volume correction to the 
splitting of a squared hadron mass, $\Delta M_{h}^2$, takes the form :
\begin{equation}\label{eq:fvol}
\Delta M_{h}^2(\infty)-\Delta M_{h}^2(L)=e^2\frac{A}{L^2}
\end{equation}
where $L$ is the spatial extent of the lattice and $A$ is an unknown 
dimensionless constant. We found, as presented in the next section, that 
finite volume corrections of type (\ref{eq:fvol}) fit lattice data 
well but are not compatible with $\SU(3)$ PQ$\chi$PT+QED (\cf Figure 
\ref{fig:fvol}).

\section{Preliminary results}\label{sec:res}
\begin{figure}[h]
\centering
\includegraphics[width=\plots\textwidth]{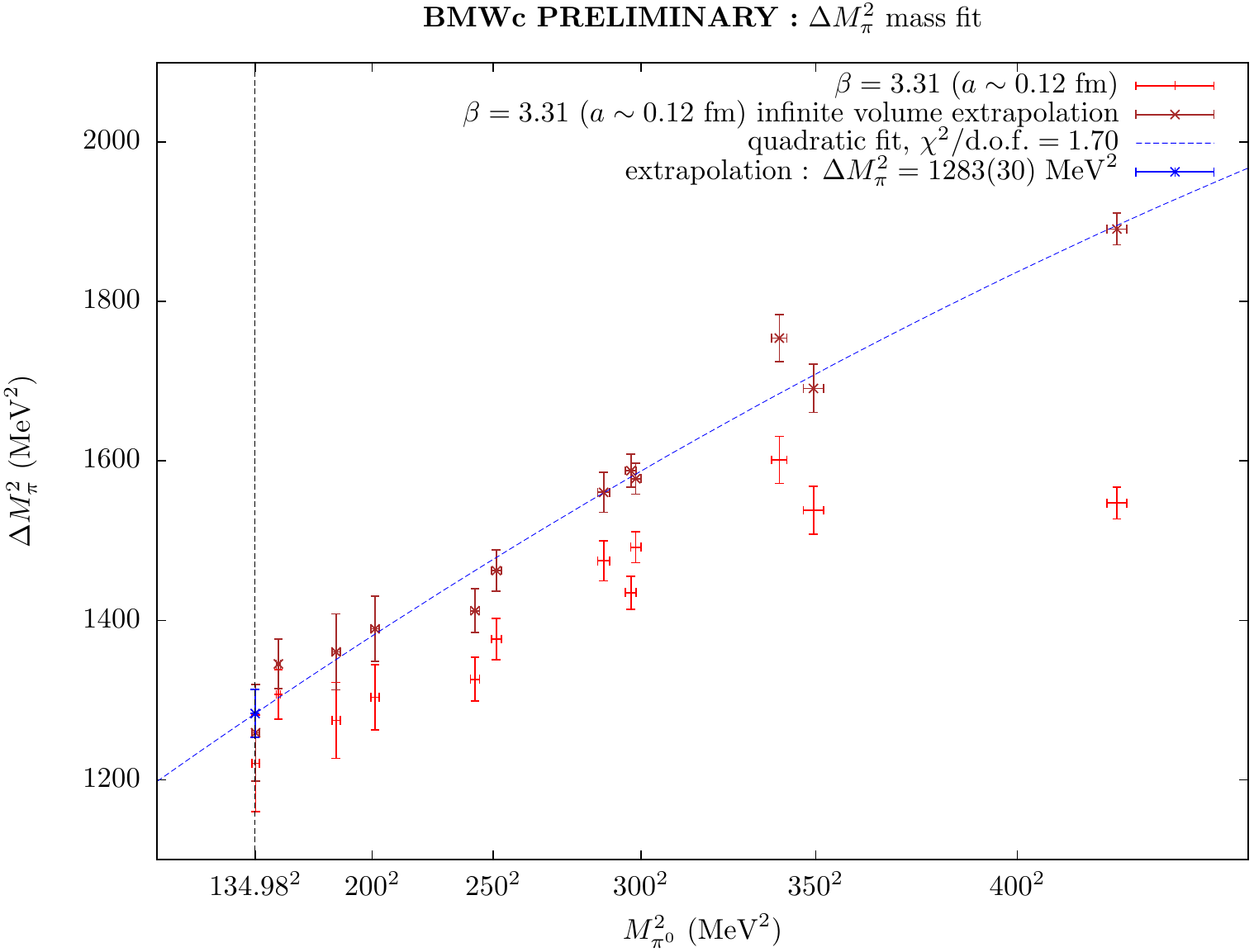}
\caption{Pion squared mass splitting \textit{vs.} neutral pion squared mass. 
Red points represent raw lattice data, dark red points are the same data with 
finite volume correction of type (\protect\ref{eq:fvol}) and the dashed blue 
line is the result of the physical point fit.\label{fig:dMsqpi}}
\end{figure}

For this preliminary analysis, we used the  $\beta=3.31$ subset of 
Budapest-Marseille-Wuppertal 2-HEX gauge configurations~\citep{Durr:2010tu}.
The main features of this dataset are:
\begin{itemize}
    \item one lattice spacing $a\simeq\unit{0.12}{\fm}$
    \item $12$ pion masses from $\unit{135}{\mega\ev}$ to $\unit{422}{\mega\ev}$
    \item $4$ spatial volumes from $(\unit{2}{\fm})^3$ to
     $(\unit{5.8}{\fm})^3$, with $M_{\pi^0}L>4$
\end{itemize}
To interpolate a quantity to the physical masses, we use a Taylor
expansion in $M_{\pi^0}^2$ and
\\ $M_{K_{\chi}}^2=\frac{1}{2}(M_{K^+}^2+M_{K^0}^2-M_{\pi^+}^2)$. We
choose $M_{\pi^0}^2$ because it has negligible finite volume
corrections compared to those in $M_{\pi^+}^2$. The lattice spacing is
computed in physical units using the $\Omega^-$ baryon mass. The
infinite volume extrapolation is performed by including corrections of
the kind (\ref{eq:fvol}) into the fit. Fits are carried out using
fully correlated $\chi^2$ minimization and a bootstrap error
analysis. Using this methodology, we obtain the following preliminary
results (an example of such a fit is shown in Figure \ref{fig:dMsqpi})
:
\begin{eqnarray}
\Delta_{\mathrm{EM}}M_{K}^{2} & = & 
	\unit{2179\err{34}{stat.}\err{100}{qu.}\err{?}{sys.}}{\mega\ev\squared}\\
\Delta_{\mathrm{EM}}M_{\pi}^{2} & = & 
	\unit{1283\err{30}{stat.}\err{0}{qu.}\err{?}{sys.}}{\mega\ev\squared}\\
\Delta_{A}D & = &
	\unit{896\err{37}{stat.}\err{100}{qu.}\err{?}{sys.}}{\mega\ev\squared}\\
\Delta_{R}D & = &
	0.70\err{4}{stat.}\err{8}{qu.}\err{?}{sys.}
\end{eqnarray}
where $\err{?}{sys.}$ stands for the systematic errors which we have
not yet estimated, such as those associated with taking the continuum
limit, etc.

\section{Conclusion}
Using the methodology that we presented last
year~\citep{Portelli:2010tz}, we extended our dataset down to the
physical value of the light quark mass. We also studied two important
systematics effects: QED quenching and finite volume effects.

Our results are promising and close to phenomenological 
estimates~\citep{Colangelo:2010uu}. Moreover, concerning the finite volume 
effects, the results are compatible with a simple $\frac{1}{L^2}$ 
model and it seems that $\SU(3)$ PQ$\chi$PT+QED fails to describe them.

In the short term, we will continue our analysis on several lattice
spacings with non-degenerate up and down quark masses. This will
provide the last ingredients to define precisely the physical point,
which require a continuum limit and physical isospin breaking.

\acknowledgments{Computations were performed using HPC resources from FZ 
Jülich and from GENCI-[IDRIS/CCRT] (grant 52275) and clusters at Wuppertal 
and CPT. This work is supported in part by EU grants I3HP, FP7/2007-2013/ERC 
No. 208740, MRTN-CT-2006-035482 (FLAVIAnet), DFG grants FO 502/2, SFB-TR 55, 
by CNRS grants GDR 2921 and PICS 4707.}

\bibliographystyle{mybib}
\bibliography{article}

\begin{thebibliography}{19}
\newcommand{\enquote}[1]{``#1''}
\providecommand{\natexlab}[1]{#1}
\providecommand{\url}[1]{\texttt{#1}}
\providecommand{\urlprefix}{URL }
\providecommand{\eprint}[2][]{\url{#2}}

\bibitem[{D{\"u}rr \emph{et~al.}(2011)}]{Durr:2010uf}
S.~D{\"u}rr \emph{et~al.}
\newblock \enquote{{Lattice QCD at the physical point: light quark masses}.}
\newblock In \emph{Physics Letters B}, 701(2):265--268 (2011).

\bibitem[{Dürr \emph{et~al.}(2011)}]{Durr:2010tu}
S.~Dürr \emph{et~al.}
\newblock \enquote{{Lattice QCD at the physical point: Simulation and analysis
  details}.}
\newblock In \emph{JHEP}, (8):1--47 (2011).

\bibitem[{Colangelo \emph{et~al.}(2011)}]{Colangelo:2010uu}
G.~Colangelo \emph{et~al.}
\newblock \enquote{{Review of lattice results concerning low energy particle
  physics}.}
\newblock In  (2011).
\newblock \href{http://arxiv.org/abs/1011.4408v1}{\mbox{hep-lat/1011.4408}}.

\bibitem[{Hoelbling(2011)}]{Hoelbling:2011kk}
C.~Hoelbling.
\newblock \enquote{{Light hadron spectroscopy and pseudoscalar decay
  constants}.}
\newblock In \emph{PoS (Lattice 2010)} (2011).
\newblock \href{http://arxiv.org/abs/1102.0410}{\mbox{hep-lat/1102.0410}}.

\bibitem[{Mawhinney(2010)}]{Mawhinney:2011}
R.~Mawhinney.
\newblock \enquote{{Direct and Indirect Kaon Physics Directly Below KT-22: A
  Lattice 2011 Review}.}
\newblock In \emph{PoS (Lattice 2011)} (2010).
\newblock To be published.

\bibitem[{Dashen(1969)}]{Dashen:1969tn}
R.~Dashen.
\newblock \enquote{{Chiral $\mathrm{SU}(3)\otimes\mathrm{SU}(3)$ as a symmetry
  of the strong interactions}.}
\newblock In \emph{Physical Review}, 183(5):1245--1260 (1969).

\bibitem[{Donoghue \emph{et~al.}(1993)Donoghue, Holstein, and
  Wyler}]{Donoghue:1993bm}
J.~F. Donoghue, B.~R. Holstein, and D.~Wyler.
\newblock \enquote{{Electromagnetic self-energies of pseudoscalar mesons and
  Dashen's theorem}.}
\newblock In \emph{Physical Review D}, 47(5):2089--2097 (1993).

\bibitem[{Bijnens(1993)}]{Bijnens:1993go}
J.~Bijnens.
\newblock \enquote{{Violations of Dashen's theorem}.}
\newblock In \emph{Physics Letters B}, 306(3-4):343--349 (1993).

\bibitem[{Baur and Urech(1996)}]{Baur:1996gf}
R.~Baur and R.~Urech.
\newblock \enquote{{Corrections to Dashen's theorem}.}
\newblock In \emph{Physical Review D}, 53(11):6552--6557 (1996).

\bibitem[{Bijnens and Prades(1997)}]{Bijnens:1997ku}
J.~Bijnens and J.~Prades.
\newblock \enquote{{Electromagnetic corrections for pions and kaons: Masses and
  polarizabilities}.}
\newblock In \emph{Nuclear Physics B}, 490(1-2):239--271 (1997).

\bibitem[{Gao \emph{et~al.}(1997)Gao, Li, and Yan}]{Gao:1997hq}
D.-N. Gao, B.~A. Li, and M.-L. Yan.
\newblock \enquote{{Electromagnetic mass splittings of $\pi$, $a_{1}$, $K$,
  $K_{1}(1400)$, and $K^{*}(892)$}.}
\newblock In \emph{Physical Review D}, 56(7):4115--4132 (1997).

\bibitem[{Bijnens and Danielsson(2007)}]{Bijnens:2007jk}
J.~Bijnens and N.~Danielsson.
\newblock \enquote{{Electromagnetic corrections in partially quenched chiral
  perturbation theory}.}
\newblock In \emph{Physical Review D}, 75(1):014505 (2007).

\bibitem[{Duncan \emph{et~al.}(1996)Duncan, Eichten, and
  Thacker}]{Duncan:1996cy}
A.~Duncan, E.~Eichten, and H.~Thacker.
\newblock \enquote{{Electromagnetic splittings and light quark masses in
  lattice QCD}.}
\newblock In \emph{Physical Review Letters}, 76(21):3894--3897 (1996).

\bibitem[{Blum \emph{et~al.}(2007)}]{Blum:2007gh}
T.~Blum \emph{et~al.}
\newblock \enquote{{Determination of light quark masses from the
  electromagnetic splitting of pseudoscalar meson masses computed with two
  flavors of domain wall fermions}.}
\newblock In \emph{Physical Review D}, 76(11):114508 (2007).

\bibitem[{Basak \emph{et~al.}(2008)}]{Basak:2008td}
S.~Basak \emph{et~al.}
\newblock \enquote{{Electromagnetic splittings of hadrons from improved
  staggered quarks in full QCD}.}
\newblock In \emph{PoS (Lattice 2008)} (2008).
\newblock \href{http://arxiv.org/abs/0812.4486v1}{\mbox{hep-lat/0812.4486}}.

\bibitem[{Portelli \emph{et~al.}(2010)}]{Portelli:2010tz}
A.~Portelli \emph{et~al.}
\newblock \enquote{{Electromagnetic corrections to light hadron masses}.}
\newblock In \emph{PoS (Lattice 2010)} (2010).
\newblock \href{http://arxiv.org/abs/1011.4189}{\mbox{hep-lat/1011.4189}}.

\bibitem[{Blum \emph{et~al.}(2010)}]{Blum:2010ym}
T.~Blum \emph{et~al.}
\newblock \enquote{{Electromagnetic mass splittings of the low lying hadrons
  and quark masses from 2+1 flavor lattice QCD+QED}.}
\newblock In \emph{Physical Review D}, 82:094508 (2010).

\bibitem[{D\"urr \emph{et~al.}(2011)}]{Durr:2011mp}
S.~D\"urr \emph{et~al.}
\newblock \enquote{{Precision computation of the kaon bag parameter}.}
\newblock In \emph{Physics Letters B}, 705(5):477--481 (2011).

\bibitem[{Hayakawa and Uno(2008)}]{Hayakawa:2008ci}
M.~Hayakawa and S.~Uno.
\newblock \enquote{{QED in finite volume and finite size scaling effect on
  electromagnetic properties of hadrons}.}
\newblock In  (2008).
\newblock \href{http://arxiv.org/abs/0804.2044}{\mbox{hep-lat/0804.2044}}.

\end{thebibliography}

\end{document}